\documentclass[sigconf]{acmart}
\usepackage{amsmath,amsfonts, amsthm}
\usepackage{algorithmic}
\usepackage{algorithm}
\usepackage{array}
\usepackage[caption=false,font=normalsize,labelfont=sf,textfont=sf]{subfig}
\usepackage{textcomp}
\usepackage{stfloats}
\usepackage{url}
\usepackage{verbatim}
\usepackage{graphicx}
\usepackage{tikz}
\usepackage[capitalize]{cleveref}

\AtBeginDocument{%
  }

\copyrightyear{2025}
\acmYear{2025}
\setcopyright{cc}
\setcctype{by}
\acmConference[HSCC '25]{28th ACM International Conference on Hybrid Systems: Computation and Control}{May 6--9, 2025}{Irvine, CA, USA}
\acmBooktitle{28th ACM International Conference on Hybrid Systems: Computation and Control (HSCC '25), May 6--9, 2025, Irvine, CA, USA}
\acmDOI{10.1145/3716863.3718054}
\acmISBN{979-8-4007-1504-4/2025/05}

\begin{document}

\title{Robust Aggregation of Electric Vehicle Flexibility}


\author{Karan Mukhi}
\affiliation{%
  \institution{University of Oxford}
  \city{Oxford}
  \country{United Kingdom}}
\email{karan.mukhi@cs.ox.ac.uk}

\author{Chengrui Qu}
\affiliation{%
  \institution{Peking University}
  \city{Beijing}
  \country{China}}
\email{qcr2021@stu.pku.edu.cn}

\author{Pengcheng You}
\affiliation{%
  \institution{Peking University}
  \city{Beijing}
  \country{China}}
\email{pcyou@pku.edu.cn}

\author{Alessandro Abate}
\affiliation{%
  \institution{University of Oxford}
  \city{Oxford}
  \country{United Kingdom}}
\email{alessandro.abate@cs.ox.ac.uk}

\renewcommand{\shortauthors}{Mukhi et al.}

\begin{abstract}
  We address the problem of characterizing the aggregate flexibility in populations of electric vehicles (EVs) with uncertain charging requirements. 
Extending upon prior results that provide exact characterizations of aggregate flexibility in populations of electric vehicle (EVs), we adapt the framework to encompass more general charging requirements. In doing so we give a characterization of the exact aggregate flexibility as a \textit{generalized polymatroid}. Furthermore, this paper advances these aggregation methodologies to address the case in which charging requirements are uncertain. In this extended framework, requirements are instead sampled from a specified distribution. In particular, we construct \textit{robust aggregate flexibility sets}, sets of aggregate charging profiles over which we can provide probabilistic guarantees that actual realized populations will be able to track. By leveraging measure concentration results that establish powerful finite sample guarantees, we are able to give tight bounds on these robust flexibility sets, even in low sample regimes that are well suited for aggregating small populations of EVs. We detail explicit methods to tractably compute these sets. Finally, we provide numerical results that validate our results and case studies that demonstrate the applicability of the theory developed herein. 
\end{abstract}

\ccsdesc[500]{Hardware~Safety critical systems}
\ccsdesc[500]{Hardware~Smart grid}
\ccsdesc[300]{Hardware~Batteries}
\ccsdesc[500]{Computing methodologies~Planning under uncertainty}
\ccsdesc[300]{Mathematics of computing~Combinatorial optimization}

\keywords{Smart Grid, Aggregate Flexibility, Minkowski Sum, Distributionally Robust Optimization, Uncertainty Quantification}

\maketitle
\section{Introduction}
Increasing penetrations of intermittent renewable generation are requiring power system operators to procure large amounts of flexibility in order to mitigate their variability.
Furthermore, as demand in the distribution networks grows, system operators are beginning to rely on these flexible loads to alleviate congestion in their networks, in particular by implementing local flexibility markets. 
Alongside this, EV uptake is rising and will make up a substantial portion of system demand by the end of the decade.
As they are typically plugged in for more time than they require to charge, they inherently possess a degree of \textit{flexibility} in their charging behavior. Formally, there are a set of charging profiles that an EV may take, whilst satisfying the charging requirements defined by its user. Under certain charging models, this \textit{flexibility set} can be represented as a convex polytope \cite{Zhao2016ExtractingApproximation}. 
By controlling their charging profiles, populations of EVs present a large potential source of flexibility to power systems.
However, owing to complexity and reliability constraints it is not viable for individual loads to participate in the flexibility markets, and so hierarchical control architectures have been proposed \cite{Callaway2011AchievingLoads}.
From this, aggregators, entities that collate flexibility from individual devices and bid it into the markets, have emerged.
In order to participate in these markets, aggregators must represent the aggregate flexibility in the populations of devices they control. A growing trend in the literature has been to characterize the aggregate flexibility by computing the Minkowski sum of the individual flexibility sets of devices in the population. As calculating the Minkowski sum is NP-hard, most of the work in the literature focuses on computing inner or outer \cite{Barot2017APolytopes} approximations of the aggregate flexibility sets. Inner approximations are more common as they guarantee the feasibility of all aggregate charging profiles contained within them. 
Zonotopes are used to approximate individual flexibility sets in \cite{Muller2019AggregationResources}, from which Minkowski sums can be computed efficiently. Similar to this \cite{Hao2015AggregateLoads} and \cite{Taha2024AnPopulations} use homothets instead of zonotopes. 
In \cite{Nazir2018InnerResources} a union-based approach is developed that can produce tight approximations, at the expense of added computational burden. As they do not contain all feasible aggregate charging profiles, inner approximations do not guarantee optimality.
Whilst there have been exact characterizations of the aggregate flexibility in populations of EV such as \cite{Trangbaek2012ExactDistribution} and \cite{Wen2022AggregateModels}, these methods fail to scale well in the length of the time horizon.
Thus far, only \cite{Mukhi2023AnVehicles} and \cite{Panda2024EfficientVehicles} offer exact characterizations that are computationally tractable for long time horizons. However, to scale well these works require a certain amount of homogeneity in the populations they aggregate.

\begin{figure*}[t!]
    \centering
    \includegraphics[width=0.95\textwidth]{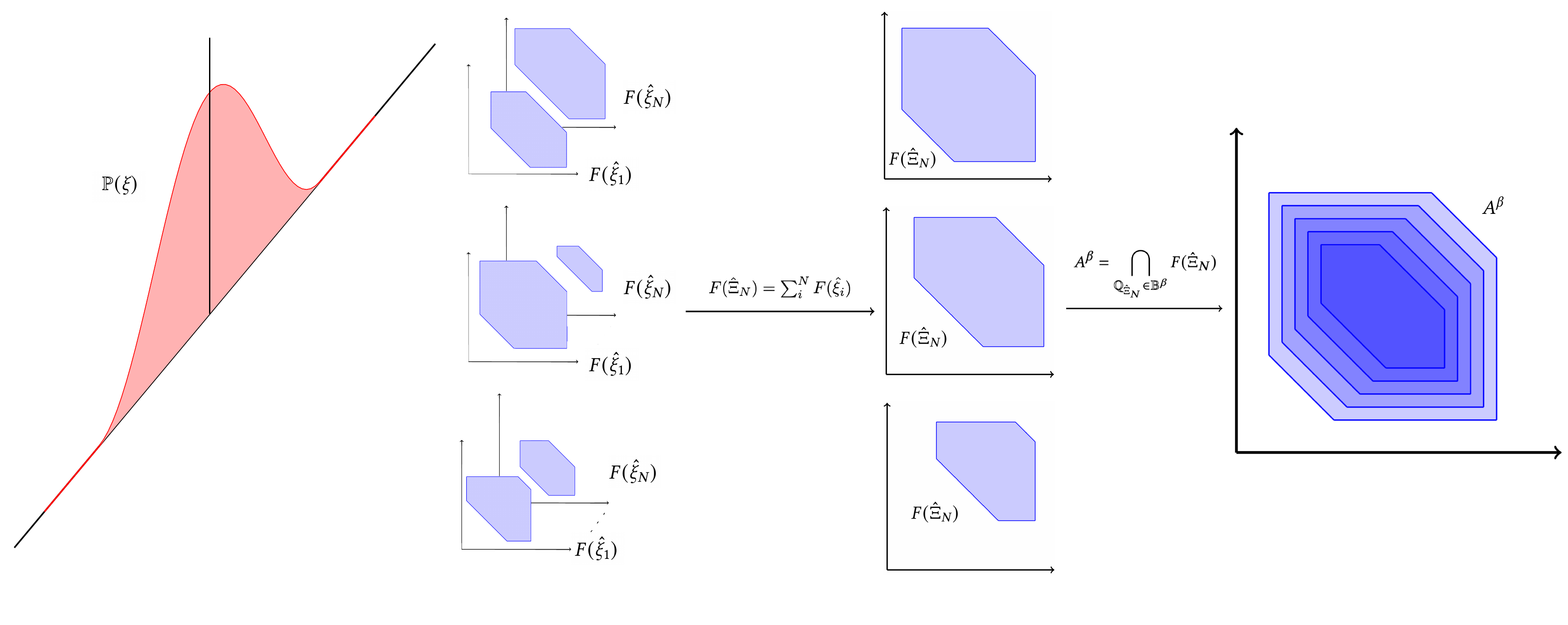}
    \caption{A schematic of the work presented in this paper. We are provided with a distribution over charging requirements $\mathbb{P}$. From this distribution, a population of $N$ charging requirements is obtained by drawing $N$ independent samples from $\mathbb{P}$. Each of the charging requirements generate their own individual flexibility sets $F(\hat{\xi}_i)$. The aggregate flexibility set for the population is the Minkowski sum of these, $F(\hat{\Xi}_i) = \sum_i^N F(\hat{\xi}_i)$, characterizing this is the subject of \cref{sec:aggregation}. This aggregate flexibility set is a random object and so we would like define robust sets $A^\beta$, in which we have varying confidence that $F(\hat{\Xi}_i)$ will be contained in. We derive these robust sets in \cref{sec:agg_uncertainty} and provide a tractable reformulation for their computation in \cref{sec:reformulation}.}
    \label{fig:schematic}
    \Description{Schematic of this work}
  \end{figure*}

All of these methods assume full knowledge of the charging requirements of the EVs they aggregate. However, in most realistic scenarios the charging requirement of an EV will be unknown to the aggregator before it plugs in, and so aggregators will need to estimate the aggregate flexibility sets. Historical datasets of EV charging requirements are often available to aggregators, such as \cite{Lee2019ACN-Data:Dataset}, and can be used to inform these estimates.
To this end, aggregators require techniques to characterize aggregate flexibility sets from historic data and uncertainty quantification methods for quantifying the confidence they have in these sets. In some scenarios, aggregators will be aggregating large populations, in which case uncertainties can largely be ignored.
However, with the advent of local flexibility markets aggregators will have to characterize the flexibility in small populations of EVs, where there are uncertainties \cite{Lampropoulos2019ACompanies}. 
In \cite{Gade2024LeveragingMarkets} the problem of bidding into energy markets with minimum reliability requirements is considered. Similar to the work presented in this paper, the authors address this by formulating it as a distributionally robust joint chance-constrained optimization problem, however they only characterize availability of populations of EVs rather than the population's flexibility over time.
Building on this, \cite{Zhou2024AggregatedGuarantee} introduces a novel approach for deriving a distributionally robust inner approximation of the flexibility sets for distributed energy resources.
While effective for general DERs, the focus on network-wide integration differs from the individual device-based flexibility modeling predominant in EV-specific contexts.  In \cite{Taheri2022Data-DrivenModels}, the authors approximate the aggregate flexibility set with an ellipsoid, which is built upon a convex quadratic classifier trained on a historic dataset. This data-driven approach provides an efficient method of calculating an approximated aggregate flexibility set, however the methods here do not establish any theoretical probabilistic guarantees. In \cite{Zhang2024AUncertainty}, the uncertainties of EVs are also modeled under a distributionally robust joint chance-constrained programming framework. However, the uncertainty sets constructed could be improved as they are derived from inner approximations of the Minkowski sum.

Given these limitations on the exactness of approximations and the rigor of uncertainty modeling, this paper makes the following contributions:
\begin{itemize}
    \item We derive a novel, efficient method of computing the aggregate flexibility set for large populations of EVs over long time horizons by expressing each EV’s feasible charging region as a generalized polymatroid. This family of polytopes is closed under Minkowski summation. Leveraging this property we obtain an exact representation of the aggregate flexibility. Unlike previously proposed methods, our approach does not resort to approximations, thus preserving all feasible solutions and soundness of the characterization.
    \item Building on this exact characterization, we exploit measure concentration results to bound the uncertainty in EV charging requirements. Crucially, these results remain tight even in low-sample regimes, making them highly applicable to scenarios involving smaller EV populations or limited historical data. Our approach thus yields distributionally robust flexibility sets, ensuring high-confidence feasibility without over constraining the available flexibility.
\end{itemize}

\cref{fig:schematic} offers a high-level schematic for the work presented here. The rest of the paper is structured as follows: \cref{sec:problem_formulation} formally introduces our setting, including the model for EV charging, the concept of individual and aggregate flexibility, and the uncertainty assumptions. In \cref{sec:aggregation} we introduce generalized polymatroids, a family of polytopes that contain all individual EV flexibility sets. We use properties of these polytopes to provide a fast aggregation method. Moving to the stochastic setting, in \cref{sec:agg_uncertainty}, we establish probabilistic guarantees built on measure concentration results, and use them to derive conditions on the parameters that define the robust flexibility sets. 
\cref{sec:reformulation} offers a derivation of how to compute the parameters defining these robust flexibility sets. We reformulate the associated optimization problems, demonstrating how to handle distributionally robust considerations efficiently.
Numerical simulations are used to validate the theoretical results of this paper in \cref{sec:case_studies}, along with a topical case study to show the practical applicability of the theory. 
Finally we discuss the implications of this work and give directions for future work in \cref{sec:conc}.

\subsection*{Notation}
We let calligraphic letters define finite sets, e.g. $\mathcal{T} = \{0,..,T-1\}$. For a vector $x \in \mathbb{R}^T$ and $t \in \mathcal{T}$, we let $x(t)$ denote the $t$-th element of $x$.
For a set $\mathcal{A} \subseteq \mathcal{T}$, we let $x(\mathcal{A}) = \sum_{t \in \mathcal{A}} x(t)$, and $x(\mathcal{A}) = 0$ for $\mathcal{A} = \emptyset$.
Uppercase letters denote convex polytopes e.g. $F \subset \mathbb{R}^T$, bold face letters, $\mathbb{P}$, denote probability distributions, and hatted letters, $\hat{\xi}$, denote random variables. $\sum$ denotes both sums and Minkowski sums, depending on the context.
\section{Problem Formulation}\label{sec:problem_formulation}
In this section, we first introduce our model for EV charging and formalize our notion of \textit{flexibility}, both in the context of individual EVs and for aggregations of them. We then introduce the problem of robust aggregation. In the following, we consider an aggregator that has direct control over the charging of a population of $N$ EVs. We denote the set of EVs as $i \in \mathcal{N} = \{1, \ldots, N\}$. We will consider the problem over a finite time horizon, and we discretize this horizon into $T$ time steps each of duration $\delta$. We denote the set of time steps as $t \in \mathcal{T} = \{0, \ldots, T-1\}$.

\subsection{Individual Flexibility Sets}
Let $u: \mathbb{R} \rightarrow \mathbb{R}$ denote the \textit{charging profile} of an EV, such that $u(\tau)$ is the power consumption at time $\tau$.
We assume that the EV's power consumption is constant in each timestep so that $u$ is piecewise constant, i.e. $u: \mathcal T  \rightarrow \mathbb{R}$, and we instead write $u \in \mathbb{R}^T$. 
We let $x \in \mathbb{R}^T$ denote the \textit{state of charge} of the EV in each time step, such that $x(t)$ is the state of charge in timestep $t$.
The charging dynamics of the EV are given by:
\begin{equation}\label{eq:ev_dynamics}
    x(t+1) = x(t) + u(t) \delta.
\end{equation}
Without loss of generality, we assume $\delta=1$ henceforth. Also, by convention, we assume that the state of charge of the EV at $t=0$ is $x(0) = 0$. This assumption is also without loss of generality, as we can simply shift the state of charge constraints in \eqref{eq:energy_constraints} when this does not hold.

In this paper we restrict the EVs to a charging-only regime, i.e. we consider no vehicle-to-grid capabilities. The maximum rate at which the EV may charge is denoted $m$. 
An EV will arrive at and depart from the charging station within the time horizon.
We denote its arrival and departure times as $\underline{t}, \overline{t} \in [0,T)$, respectively, and let $\mathcal{C} := [\underline{t}, \overline{t}] \cap \mathcal{T} \subseteq \mathcal{T}$ denote its \textit{connection time}, namely the finite set of time steps in which the EV is connected to the charging station.
At all time steps during its connection time, the EV may charge up to its maximum rate, whereas for all other time steps the EV does not charge:
\begin{subequations}\label{eq:power_constraints}
    \begin{align}
        0 \leq &u(t) \leq m \quad \forall t \in \mathcal{C}, \\
        & u(t) = 0 \quad \forall t \in \mathcal{T} \setminus \mathcal{C}.
    \end{align}
\end{subequations}
Finally, we let $\underline{e}$ and $\overline{e}$ denote the lower and upper bound on the EV's state of charge at the final time:
\begin{equation}\label{eq:energy_constraints}
    \underline{e} \leq u(\mathcal{T}) \leq \overline{e}.
\end{equation}
where $u(\mathcal{T})$ denotes the sum over all time steps, as introduced in the notation section. Note that \eqref{eq:power_constraints} ensures that the EV only charges during its connection times. For ease of notation we collect all the parameters relating to the EV charging requirements into the tuple $\xi = (\underline{e}, \overline{e}, \underline{t}, \overline{t}, m)$. Naturally, for some values of $\xi$, there will not exist any feasible charging profiles, and so we define $\Xi$ to be the set of charging requirements that are feasible:
\begin{equation}\label{eq:Xi_space}
    \Xi := \left\{ \xi=(\underline{e}, \overline{e}, \underline{t}, \overline{t}, m) \middle\vert 
    \begin{array}{@{}cl}
        &   0 \leq \underline{t} \leq \overline{t} < T \\
        &   0 \leq \underline{e} \leq \overline{e} \leq |\mathcal{C}| m  \\
\end{array} 
\right\}.
\end{equation}

Using this model, we consolidate the constraints of \eqref{eq:power_constraints} and \eqref{eq:energy_constraints} to define the set of charging profiles an EV may take whilst satisfying its charging requirements:
\begin{definition}
    For an EV with charging parameters  $\xi$, the \textit{individual flexibility set}, denoted, $ F(\xi) \subset \mathbb{R}^T$, is the set of all feasible charging profiles for the EV:
    \begin{equation}\label{eq:individual_flex_set}
         F(\xi) := \left\{ u \in \mathbb{R}^T \; \middle\vert \;
        \begin{array}{@{}cl}
                          &  0 \leq u(t) \leq 0  \;\;\; \forall t \in \mathcal{T} \setminus \mathcal{C} \\
                     &0 \leq  u(t) \leq m  \;\; \forall t \in \mathcal{C}  \\
             &\underline{e} \leq u(\mathcal{T})\leq \overline{e}  \\
        \end{array} 
        \right\}.
    \end{equation}
\end{definition}
The individual flexibility set, as defined above, is the intersection of a collection of half-spaces and so is a convex polytope in $\mathbb{R}^T$.

\subsection{Aggregate Population Flexibility Sets}
We now consider the case where we have a population of $N$ EVs, each with charging parameters $\xi_i$ for $i \in \mathcal{N}$, where $u_i$ denotes the charging profile of the $i^{th}$ EV. We let $\Xi_N$ denote the set of the charging parameters of all EVs in the population such that $\Xi_N = \{\xi_1,\ldots, \xi_N\}$.

The \textit{aggregate charging profile}, $u_{\mathcal{N}}$, of the population is given by the sum of the individual charging profiles of EVs in the population:
\begin{equation*}
    u_{\mathcal{N}} = \sum_{i=1}^N u_i.
\end{equation*}
\begin{definition}
    For a population of EVs with charging parameters, $\Xi_N$, the \textit{aggregate flexibility set}, denoted $ F(\Xi_N)$, is the set of all feasible aggregate charging profiles of the population:
    \begin{equation}\label{eq:aggregate_flex_set}
         F(\Xi_N) := \left\{ u_{\mathcal{N}} \in \mathbb{R}^T \mid u_{\mathcal{N}} = \sum_{k=1}^N u_i,\; u_i \in  F(\xi_i)\;\; \forall i\right\}.
    \end{equation}
\end{definition}
By definition, $ F(\Xi_N)$ is the Minkowski sum of the individual flexibility sets of the EVs in the population:
\begin{equation}\label{eq:mink_sum}  
     F(\Xi_N) = \sum_{i=1}^N  F(\xi_i).
\end{equation}
In general, however, an aggregator will have to make decisions about the population of EVs before they connect, and so will not have access to the charging requirements of the EVs in the population. However, they will have access to historic data on the charging requirements of the EVs. 
Therefore, we model the charging requirements as i.i.d. random variables according to a given distribution  $\mathbb{P}$ that is supported on $\Xi$. 

We denote the set of random charging requirements of the population as $\hat{\Xi}_N:= \{\hat{\xi}_i\}_{i\leq N} \subseteq \Xi$: this set is a random variable distributed according to $\mathbb{P}^N$. 
Hence, the polytope, $ F(\hat{\Xi}_N)$, generated by $\hat{\Xi}_N$, is a random object that is governed by the distribution $\mathbb{P}^N$. The aggregator will need to optimize over $F(\hat{\Xi}_N)$, before observing $\hat{\Xi}_N$. Therefore it is useful for aggregators to understand which aggregate charging profiles can be satisfied by the population within certain confidence levels. Characterizing sets of this type is the main focus of this paper.
\begin{definition}\label{dfn:robust_flex}
    For a population of $N$ EVs with charging requirements $\hat{\Xi}_N$, distributed according to $\mathbb{P}^N$, the \textit{robust aggregate flexibility set} at confidence level $\beta$, denoted $A_{\mathbb{P}, N}^\beta$, is the set of all aggregate charging profiles that can be satisfied with a probability of $1 - \beta$:
    \begin{equation*}
        A_{\mathbb{P}, N}^\beta := \{u_{\mathcal{N}} \in \mathbb{R}^T \; | \; \mathbb{P}^N\{\hat{\Xi}_N: u_{\mathcal{N}} \in  F(\hat{\Xi}_N)\} \geq 1 - \beta \}.
    \end{equation*}
\end{definition}
Here, we emphasize that $ A_{\mathbb{P}, N}^\beta$ represents a set of aggregate charging profiles for a population of $N$ electric vehicles, each with charging requirements sampled from distribution $\mathbb{P}$, for which we can guarantee will be feasible with a confidence level of $1 - \beta$. In the following we assume the distribution over charging requirements $\mathbb{P}$ and the size of the population $N$ is fixed, and so, for clarity, we will drop the superscripts $\mathbb{P}$ and $N$ from $A_{\mathbb{P}, N}^\beta$, and simply refer to it as $A^\beta$.
\subsection{Limitations}
Although the proposed model offers a realistic treatment of EV charging requirements, it is subject to some limitations. Firstly, we assume that the EV charging requirements are i.i.d. according to the common distribution $\mathbb{P}$. This assumption may hold in some contexts, for example, in a charging station serving a homogeneous fleet. However, in scenarios where aggregators have built up a profile of individual vehicles in the population, it becomes more appropriate to 
model each with distinct distributions.
In such cases, the i.i.d. assumption becomes limiting and may overlook important heterogeneity.

Secondly, our model restricts EVs to charging only, with no allowance for discharging. Generalized polymatroids are expressive enough to faithfully model discharging. However, this adds another layer of complexity to the characterizations. The focus of this paper is on uncertainty quantification in the aggregate flexibility sets, and so we work with this simpler, but less expressive, charging model. Applying this work to the more general case of EVs with discharging capabilities is an interesting avenue for future work.
\section{Aggregation of Flexibility}\label{sec:aggregation}
In this section, we provide a characterization of the aggregate flexibility set defined in \eqref{eq:aggregate_flex_set}. We obtain this by computing the Minkowski sum of the individual flexibility sets. In general, computing Minkowski sums is an NP-hard problem, however in this section we provide a tractable method for its computation. This is done by showing that the individual flexibility sets, as defined in \eqref{eq:individual_flex_set}, are members of a family of polytopes that are closed under Minkowski summation, and whose Minkowski sum can be computed efficiently.

\begin{figure}[t!]
    \centering
    \begin{tikzpicture}
    \filldraw[opacity=0.3, blue] (1,3) -- (3,1) -- (6,1) -- (6,6) -- (1,6) -- cycle;
    \filldraw[opacity=0.3, red] (0,0) -- (4.5,0) -- (4.5,3.5) -- (3.5,4.5) -- (0,4.5) -- cycle;
    \draw[-][line width=0.4mm][blue] (1,3) -- (3,1);
    \draw[-][line width=0.1mm][black] (1,3) -- (1,6);
    \draw[-][line width=0.1mm][black] (6,1) -- (3,1);
    \draw[-][line width=0.3mm][red] (4.5,3.5) -- (3.5,4.5);
    \draw[-][line width=0.1mm][black] (4.5,3.5) -- (4.5,0);
    \draw[-][line width=0.1mm][black] (3.5,4.5) -- (0,4.5);
    \draw[-][line width=0.4mm, dashed][black] (4.5,3.5) -- (4.5,1);
    \draw[-][line width=0.4mm, dashed][black] (3.5,4.5) -- (1,4.5);
    \draw[-][line width=0.4mm, dashed][black] (3,1) -- (4.5,1);
    \draw[-][line width=0.4mm, dashed][black] (1,3) -- (1,4.5);
    \coordinate (A) at (1,3);
    \coordinate (B) at (3,1);
    \filldraw (A) circle (1.5pt) node[above right] {};
    \filldraw (B) circle (1.5pt) node[above right] {};
    \coordinate (Q) at (4.5,3.5);
    \coordinate (R) at (3.5,4.5);
    \filldraw (Q) circle (1.5pt) node[above right] {};
    \filldraw (R) circle (1.5pt) node[above right] {};
    \node at (1.2,1.2) {$S(b)$};
    \node at (4.7,4.7) {$S'(p)$};
    \node at (3,3) {$Q(p, b)$};
\end{tikzpicture}
    \caption{Polyhedra generated by the paramodular pair $(p,b)$. The red and blue lines are the submodular and supermodular base polyhedra. The red and blue shaded regions are the submodular and supermodular polyhedra, $S(b)$ and $S'(p)$.  The g-polymatroid, $Q(p, b)$, is generated by the intersection of $S(b)$ and $S'(p)$.}
    \label{fig:sub_super_polyhedra}
    \Description{Submodular and supermodular polyhedra, and the g-polymatroid generated by their intersection.}
\end{figure}
We start by introducing some concepts that will be relevant for our aggregation method. In the following we will consider set functions defined over the set $\mathcal{T} = \{0,...,T-1\}$ , where $T$ is the time horizon of the problem.
\begin{definition}[Submodular functions]\cite{Frank2011ConnectionsOptimization}
    A \textit{submodular function} $b:2^\mathcal{T}\rightarrow \mathbb{R}$, is a set-function defined over the subsets of a finite set $\mathcal{T}$, which satisfies the inequality 
    \begin{equation}\label{eq:submodular}
        b(\mathcal{A} \cup \{e\}) - b(\mathcal{A}) \geq b(\mathcal{B} \cup \{e\}) - b(\mathcal{B}). 
    \end{equation}
     for all subsets $\mathcal{A}\subseteq \mathcal{B} \subseteq \mathcal{T}$ and $e \in \mathcal{T}\setminus \mathcal{B}$.
\end{definition}
One can also define a \textit{supermodular function} by reversing the inequality in \eqref{eq:submodular}, or as the negative of a submodular function: if $b$ is submodular then $p = -b$ is supermodular.
The \textit{submodular polyhedron} associated with a submodular function, $b$, is defined as:
\begin{equation*}
    S(b) := \{x \in \mathbb{R}^T\; | \; x(\mathcal{A}) \leq b(\mathcal{A}), \; \forall \mathcal{A} \subseteq \mathcal{T}\}.
\end{equation*}
The intersection of this with the plane $x(\mathcal{T}) = b(\mathcal{T})$ is called the \textit{base polyhedron} of $b$:
\begin{equation*}
    B(b) := \{x \in \mathbb{R}^T\; | \; x(\mathcal{A}) \leq b(\mathcal{A}), \; \forall \mathcal{A} \subseteq \mathcal{T}, x(\mathcal{T}) = b(\mathcal{T})\}.
\end{equation*}
Similar definitions can be made for supermodular functions, where the supermodular functions bound the polyhedra from below.
\begin{definition}[Paramodularity]\cite{Frank2011ConnectionsOptimization}
    A pair $(p,b)$ of set-functions is \textit{paramodular} if $p(\emptyset) = b(\emptyset) = 0$, $p$ is supermodular, $b$ is submodular, and for all $\mathcal{A}, \mathcal{B} \subseteq \mathcal{T}$:
    \begin{equation*}
        b(\mathcal{A}) - p(\mathcal{B}) \geq b(\mathcal{A}\setminus \mathcal{B}) - p(\mathcal{B} \setminus \mathcal{A}).
    \end{equation*}
\end{definition}
The intuition behind paramodularity is that it restricts the base polyhedra of the submodular function to be contained within the supermodular polyhedra of the supermodular function, and vice versa.
\begin{definition}[Generalized polymatroids]\cite{Frank2011ConnectionsOptimization}
    For a paramodular pair $(p,b)$, the polyhedron $Q(p,b)$ is called a \textit{generalized polymatroid} (\textit{g-polymatroid} for short), where: 
    \begin{equation*}
        Q(p,b) := \{x \in \mathbb{R}^T\; | \; p(\mathcal{A}) \leq x(\mathcal{A}) \leq b(\mathcal{A}) \; \forall \mathcal{A} \subseteq \mathcal{T}\}.
    \end{equation*}
\end{definition}
Essentially, a g-polymatroid is the intersection of a supermodular polyhedron and a submodular polyhedron, generated by a paramodular pair. A visual depiction of this construction is provided in \cref{fig:sub_super_polyhedra}.

\begin{definition}[Plank]\cite{Frank2011ConnectionsOptimization}
    Given $\alpha, \beta \in \mathbb{R}$ with $\alpha \leq \beta$, the \textit{plank} $K(\alpha, \beta)$ is
    \begin{equation*}
        K(\alpha, \beta) := \{x \in \mathbb{R}^T\; | \; \alpha \leq x(\mathcal{T}) \leq \beta\}.
    \end{equation*}
\end{definition}

We now turn our attention to the individual flexibility sets introduced in the previous section. One can write an individual flexibility set as the intersection of a cube and a plank: $F(\xi) = F'(\xi) \cap K(\underline{e}, \overline{e})$, where:
\begin{equation}\label{eq:power_matroid}
    F'(\xi) := \left\{ u \in \mathbb{R}^T \; \middle\vert \;
    \begin{array}{@{}cl}
                      &  0 \leq u(t) \leq 0  \;\;\; \forall t \in \mathcal{T} \setminus \mathcal{C} \\
                 &0 \leq  u(t) \leq m  \;\; \forall t \in \mathcal{C}
    \end{array} 
    \right\}.
\end{equation}
\begin{lemma}\label{lem:power_matroid}
    $F'(\xi)$ is the g-polymatroid $Q(p', b')$, where:
    \begin{align*}
        &p_\xi'(\mathcal{A}) := 0, \\
        &b_\xi'(\mathcal{A}) :=|\mathcal{A} \cap \mathcal{C}|m.
    \end{align*}
    Here, $|\mathcal{A}|$ denotes the cardinality of the set $\mathcal{A}$.
\end{lemma}
\begin{proof}   
    $F'(\xi)$ is a cube of side length $m$ in the subspace $\mathbb{R}^\mathcal{C} \subseteq \mathbb{R}^T$. The submodular function that generates this cube is $b_\xi'(\mathcal{A}) =|\mathcal{A} \cap \mathcal{C}|m$.
    $F'(\xi)$ is bounded from below by $0$ for all $\mathcal{A} \subseteq \mathcal{T}$, and so $p_\xi'(\mathcal{A}) = 0$.
    The paramodularity of the pair $(p_\xi', b_\xi')$ holds trivially.
\end{proof}
We can now use the following theorem to show that the intersection of $F'(\xi)$ and $K(\underline{e}, \overline{e})$, and hence $F(\xi)$, is a g-polymatroid.
\cref{fig:plank_box} provides a graphical illustration of the theorem.



\begin{theorem}[Plank Intersection theorem]\label{thm:g_polymatroid_intersection}\cite[Theorem 14.3.13]{Frank2011ConnectionsOptimization}
    \newline
    The intersection of the g-polymatroid $Q(p', b')$, and the plank $K(\alpha, \beta)$ is a g-polymatroid $Q(p, b)$, where:
    \begin{align*}
        p(\mathcal{A}) &:= \max\{p'(\mathcal{A}), \alpha - b'(\mathcal{T}\setminus \mathcal{A})\} \\
        b(\mathcal{A}) &:= \min\{b'(\mathcal{A}), \beta - p'(\mathcal{T}\setminus \mathcal{A})\}.
    \end{align*}
\end{theorem}

\begin{figure}[t!]
    \centering
    \begin{tikzpicture}
        
    \filldraw[opacity=0.3, blue] (-.8,4.8) -- (0,5.6) -- (5.6,0) --  (4.8,-.8)-- cycle;
    \filldraw[opacity=0.3, red] (0,0) -- (4.5,0) -- (4.5,4.5) -- (0,4.5) -- cycle;

    \draw[-][line width=0.1mm][black] (4.5,4.5) -- (4.5,0);
    \draw[-][line width=0.1mm][black] (4.5,4.5) -- (0,4.5);

    \draw[-][line width=0.1mm][blue] (0,5.6) -- (5.6,0);
    \draw[-][line width=0.1mm][blue] (-.8,4.8) --(4.8,-.8);

    \draw[-][line width=0.3mm][blue] (0,4) -- (4,0);
    \draw[-][line width=0.3mm][blue] (4,0) -- (4.5,0);
    \draw[-][line width=0.3mm][blue] (4.5,0) -- (4.5,1.1);
    \draw[-][line width=0.3mm][blue] (0,4) -- (0,4.5);
    \draw[-][line width=0.3mm][blue] (0,4.5) -- (1.1,4.5);
    \draw[-][line width=0.3mm][blue] (1.1,4.5) -- (4.5,1.1);

    \node at (-.5,3) {$u(2)$};
    \node at (3,-.5) {$u(1)$};
    \node at (2.38,2.38) {$Q(p', b')$};
    \draw[->][line width=0.4mm] (0,0) -- (6,0);
    \draw[->][line width=0.4mm] (0,0) -- (0,6);
    
    \draw[-][line width=0.5mm][blue] (0,4) -- (4,0);
    \draw[-][line width=0.5mm][blue] (4,0) -- (4.5,0);
    \draw[-][line width=0.5mm][blue] (4.5,0) -- (4.5,1.1);
    \draw[-][line width=0.5mm][blue] (0,4) -- (0,4.5);
    \draw[-][line width=0.5mm][blue] (0,4.5) -- (1.1,4.5);
    \draw[-][line width=0.5mm][blue] (1.1,4.5) -- (4.5,1.1);
\end{tikzpicture}
    \caption{A cube $F'(\xi)$ (red shaded region), and a plank $K(\underline{e}, \overline{e}$) (blue shaded region), intersecting to form the g-polymatroid $Q(p',b')$ (region outlined in blue).}
    \label{fig:plank_box}
    \Description{A cube and a plank intersecting to form a g-polymatroid.}
  \end{figure}
\begin{corollary}\label{cor:individual_flex} $F(\xi)$ is the g-polymatroid $Q(p_\xi, b_\xi)$, where:
    \begin{align*}
        &p_\xi(\mathcal{A}) := \max\{0, \underline{e} - |\mathcal{C} \setminus \mathcal{A}|m\}, \\
        &b_\xi(\mathcal{A}) := \min\{|\mathcal{A} \cap \mathcal{C}|m, \overline{e}\}.
    \end{align*}
\end{corollary}
\begin{proof}
    This follows directly from \cref{thm:g_polymatroid_intersection}, \cref{lem:power_matroid} and using the identity $|(\mathcal{T} \setminus \mathcal{A}) \cap \mathcal{C}| = |\mathcal{C} \setminus \mathcal{A}|, \; \forall \mathcal{C} \subseteq \mathcal{T}$.
\end{proof}
Now, with our characterization of the individual flexibility sets as g-polymatroids we can exploit some of the properties of this family of polytopes. In particular, we can use the following theorem to efficiently compute the Minkowski sum of a set of g-polymatroids.
\begin{theorem}[Sum theorem]\label{thm:g_polymatroid_sum}\cite[Theorem 14.2.15]{Frank2011ConnectionsOptimization}
    \newline
    The Minkowski sum of a set of g-polymatroids is given by
    \begin{equation*}
        \sum_i Q(p_i, b_i) = Q\left(\sum_i p_i, \sum_i b_i\right). 
    \end{equation*}
\end{theorem}

\begin{corollary}\label{cor:aggregate_flex} The aggregate flexibility set, $ F(\Xi_N)$, is the g-polymatroid $Q(p_{\Xi_N}, b_{\Xi_N})$, where:
    \begin{align*}
        &p_{\Xi_N}(\mathcal{A}) :=  \sum_{i=1}^N p_{\xi_i}(\mathcal{A}) =  \sum_{i=1}^N \max\{0, \underline{e}_i - |\mathcal{C}_i \setminus \mathcal{A}|m_i\}, \\
        &b_{\Xi_N}(\mathcal{A}) :=  \sum_{i=1}^N b_{\xi_i}(\mathcal{A}) = \sum_{i=1}^N \min\{ |\mathcal{A} \cap \mathcal{C}_i|m_i, \overline{e}_i\}.
    \end{align*}
\end{corollary}

This result provides a tractable method of computing the aggregate flexibility set, $F(\Xi_N)$, as the Minkowski sum of the individual flexibility sets. 
The representation of $F(\Xi_N)$, characterized as a generalized polymatroid, involves an exponential number of constraints, which may appear computationally intractable to handle. Nevertheless, these polytopes are well-studied, and various optimization problems can be solved in polynomial time. For a detailed discussion the reader is referred to \cite{Frank2011ConnectionsOptimization}. 
\section{Aggregation Under Uncertainty}\label{sec:agg_uncertainty}
We now turn to the case in which charging requirements are uncertain, but i.i.d. according to a known distribution $\mathbb{P}$. In practice, we assume that aggregators have a large amount of historical data of charging requirements, from which they can construct the known $\mathbb{P}$. 
We denote the set of random charging requirements of the population as $\hat{\Xi}_N:= \{\hat{\xi}_i\}_{i\leq N} \subseteq \Xi$. This set is a random variable distributed according to $\mathbb{P}^N$. As the population charging requirements are random, the aggregate flexibility set generated by them, $F(\hat{\Xi}_N)$, will also be a random object. And so we focus on characterizing a robust set of aggregate charging profiles that can be tracked with confidence $1-\beta$, namely $A^\beta$ from \cref{dfn:robust_flex}. 

To do so, we first characterize an ambiguity set of sample populations that will be realized with a given confidence.
Let $\mathbb{Q}_{\hat{\Xi}_N}$ denote the discrete empirical probability distribution generated by the sample population $\hat{\Xi}_N$:
\begin{equation*}
    \mathbb{Q}_{\hat{\Xi}_N} := \frac{1}{N}\sum_{\hat{\xi}_i \in \hat{\Xi}_N} \delta_{\hat{\xi}_i},
\end{equation*}
where $\delta_{\hat{\xi}_i}$ is the Dirac distribution that concentrates mass on $\hat{\xi}_i$.
\begin{definition}
    The Wasserstein distance between two distributions $\mathbb{P}$ and $\mathbb{Q}$ is defined as:
    \begin{equation*}
        d_W(\mathbb{P},\mathbb{Q}) := \inf_{\Pi}  \int_{\Xi \times \Xi} ||\xi - \xi'|| \Pi(d\xi, d\xi'),
    \end{equation*}
    where $\Pi$ is a joint distribution on $\Xi \times \Xi$ with marginals $\mathbb{P}$ and $\mathbb{Q}$, and $||\xi - \xi'||$ is an arbitrary norm on $\Xi$.
\end{definition}
Intuitively, the Wasserstein distance represents the minimum \textit{work} required to move the probability mass of $\mathbb{P}$ to the distribution $\mathbb{Q}$. 
The reader is referred to \cite{Villani2003TopicsTheory} for a more complete discussion of the Wasserstein distance.
With these definitions we can tap into the following result.  

\begin{theorem}[Measure concentration]\label{thm:measure_concentration} \cite[Theorem 2]{Fournier2015OnMeasure}:
\begin{equation*}
    \mathbb{P}^N \left\{\hat{\Xi}_N : d_W(\mathbb{P}, \mathbb{Q}_{\hat{\Xi}_N}) \geq \varepsilon \right\} \leq \beta
\end{equation*}
where for all $N \geq 1$

\begin{equation}\label{eq:beta_epsilon}
    \beta =   \begin{cases}
        c_1  e^{-c_2 N \varepsilon^2} & \text{if } \varepsilon \leq 1 \\[10pt]
        c_1 e^{-c_2 N \varepsilon^a} & \text{if } \varepsilon > 1
        \end{cases}
\end{equation}
and $c_1, c_2$ and $a$ are positive constants that depend on $\mathbb{P}$ and the norm used to define the Wasserstein distance.
\end{theorem}
We can now define an ambiguity set, $\mathbb{B}^\beta$, as the set of empirical distributions that lie inside a Wasserstein ball of radius $\varepsilon_N(\beta)$ centered on $\mathbb{P}$:
\begin{equation}
    \mathbb{B}^\beta := \left\{ \mathbb{Q}_{\hat{\Xi}_N} \; \middle\vert \; d_W(\mathbb{P}, \mathbb{Q}_{\hat{\Xi}_N}) \leq \varepsilon_N(\beta) \right\},
\end{equation}
where $\varepsilon_N(\beta)$ is simply given by the inverse of \eqref{eq:beta_epsilon}. With a slight abuse of notation, we will use $\hat{\Xi}_N \in \mathbb{B}^\beta$ to denote that the empirical distribution generated by the sample population $\hat{\Xi}_N$ lies within the Wasserstein ball $\mathbb{B}^\beta$, i.e.
\begin{equation}\label{eq:sample_distribution_in_wass_ball_notation}
    \hat{\Xi}_N \in \mathbb{B}^\beta \implies d_W(\mathbb{P}, \mathbb{Q}_{\hat{\Xi}_N}) \leq \varepsilon_N(\beta).
\end{equation}

Using \cref{thm:measure_concentration} we can give the following probabilistic guarantees over our ambiguity set: 
\begin{equation*}
     \mathbb{P}^N \left\{\hat{\Xi}_N : \hat{\Xi}_N \in  \mathbb{B}^\beta \right\} \geq 1 - \beta.
\end{equation*}

\subsubsection*{Remark}
This result is particularly well-suited to the problem at hand. The measure concentration result offers tight bounds when the sample size is small. In some use cases aggregators may have to optimize and coordinate the charging of EVs that are connected to the same low-voltage substation. For such cases the number of EVs in the population, $N$, is limited, making it crucial to have guarantees that perform effectively for small sample sizes.
This work can be compared to the likes of \cite{MohajerinEsfahani2018Data-drivenReformulations}, with one key distinction: we assume to have knowledge of the distribution over charging requirements $\mathbb{P}$, and access to a finite (possibly small) population that samples the distribution $\mathbb{P}$. Hence the ambiguity sets referred to in this paper are centred on $\mathbb{P}$ and we are searching for ``worst-case'' populations that generate empirical distributions, $\mathbb{Q}_{\hat{\Xi}_N}$, that lie within this set. This is opposed to \cite{MohajerinEsfahani2018Data-drivenReformulations}, which instead centers the ambiguity set on the empirical distribution of the sampled data and searches for ``worst-cases'' of the true distribution.


Using $\mathbb{B}^\beta$ as our ambiguity set, we want to ensure that for all populations that generate empirical distributions that lie in $\mathbb{B}^\beta$, the aggregate flexibility sets associated to them contain the robust aggregate flexibility set $A^\beta$. Additionally, we want to ensure that this robust aggregate flexibility set is the largest set that satisfies this property. 
\begin{lemma} The robust aggregate flexibility set can be characterized as
    \begin{equation}\label{eq:robust_intersection}
        A^\beta =  \bigcap_{ \hat{\Xi}_N \in \mathbb{B}^\beta}  F(\hat{\Xi}_N)
    \end{equation}
\end{lemma}
\begin{proof}
    From \eqref{eq:robust_intersection}, $A^\beta$ is defined as the intersection of a collection of polytopes. Consequently, $A^\beta$ is a subset of each of these polytopes, i.e.
    \begin{align*}
        A^\beta  \subseteq \;&F( \hat{\Xi}_N), \;\; \forall \;\;  \hat{\Xi}_N \in \mathbb{B}^\beta,
    \end{align*}
    and so we guarantee that $A^\beta$ is contained within all aggregate flexibility sets generated by sample populations that lie in the ambiguity set.
    Moreover, \eqref{eq:robust_intersection} defines the largest set that fulfills this property.
\end{proof}
From this implication, we can characterize $A^\beta$ by calculating the flexibility sets for all elements in our ambiguity set and taking their intersection. 
\begin{figure}[t!]
    \centering
    \begin{tikzpicture}

        \def\pOne{2.5}
        \def\pTwo{.75}
        \def\bOne{4.2}
        \def\bTwo{3.1}

        \def\pOnerobust{3.25}
        \def\pTworobust{1.8}
        \def\bOnerobust{5.5}
        \def\bTworobust{3}

        \coordinate (Arob) at (\pTworobust,\pOnerobust);
        \coordinate (Brob) at (\pTworobust,\bOne);
        \coordinate (Crob) at (\bTwo,\bOne);
        \coordinate (Drob) at (\bOne,\bTwo);
        \coordinate (Erob) at (\bOne,\pTworobust);
        \coordinate (Frob) at (\pOnerobust,\pTworobust);

        \coordinate (p1) at (\pTwo,0);
        \coordinate (p2) at (\pTwo,6);
        \coordinate (p1robust) at (\pTworobust,0);
        \coordinate (p2robust) at (\pTworobust,6);

        \coordinate (b1) at (\bOne,0);
        \coordinate (b2) at (\bOne,6);
        \coordinate (b1robust) at (\bOnerobust,0);
        \coordinate (b2robust) at (\bOnerobust,6);

        \draw[-][line width=0.2mm][red] (p1) -- (p2);
        \draw[-][line width=0.2mm][red] (p1robust) -- (p2robust);
        \draw[->][line width=0.3mm][red] (\pTwo+0.05,1) -- (\pTworobust -0.05,1);

        \draw[-][line width=0.2mm][red] (b1) -- (b2);
        \draw[-][line width=0.2mm][red] (b1robust) -- (b2robust);
        \draw[->][line width=0.3mm][red] (\bOnerobust- 0.05,1) -- (\bOne  + 0.05,1);

        \coordinate (A) at (\pTwo,\pOne);
        \coordinate (B) at (\pTwo,\bOne);
        \coordinate (C) at (\bTwo,\bOne);
        \coordinate (D) at (\bOne,\bTwo);
        \coordinate (E) at (\bOne,\pTwo);
        \coordinate (F) at (\pOne,\pTwo);
        
        \coordinate (Arobust) at (\pTworobust,\pOnerobust);
        \coordinate (Brobust) at (\pTworobust,\bOnerobust);
        \coordinate (Crobust) at (\bTworobust,\bOnerobust);
        \coordinate (Drobust) at (\bOnerobust,\bTworobust);
        \coordinate (Erobust) at (\bOnerobust,\pTworobust);
        \coordinate (Frobust) at (\pOnerobust,\pTworobust);

        \draw[-][dashed, line width=0.2mm][blue] (A) -- (B);
        \draw[-][dashed, line width=0.2mm][blue] (B) -- (C);
        \draw[-][dashed, line width=0.2mm][blue] (C) -- (D);
        \draw[-][dashed, line width=0.2mm][blue] (D) -- (E);
        \draw[-][dashed, line width=0.2mm][blue] (E) -- (F);
        \draw[-][dashed, line width=0.2mm][blue] (F) -- (A);

        \filldraw[opacity=0.2, blue] (A) -- (B) -- (C) -- (D) -- (E) -- (F) -- cycle;
        \filldraw[opacity=0.2, blue] (Arobust) -- (Brobust) -- (Crobust) -- (Drobust) -- (Erobust) -- (Frobust) -- cycle;

        \draw[-][dashed, line width=0.3mm][blue] (Arobust) -- (Brobust);
        \draw[-][dashed, line width=0.3mm][blue] (Brobust) -- (Crobust);
        \draw[-][dashed, line width=0.3mm][blue] (Crobust) -- (Drobust);
        \draw[-][dashed, line width=0.3mm][blue] (Drobust) -- (Erobust);
        \draw[-][dashed, line width=0.3mm][blue] (Erobust) -- (Frobust);
        \draw[-][dashed, line width=0.3mm][blue] (Frobust) -- (Arobust);

        \draw[-][line width=0.3mm][blue] (Arob) -- (Brob);
        \draw[-][line width=0.3mm][blue] (Brob) -- (Crob);
        \draw[-][line width=0.3mm][blue] (Crob) -- (Drob);
        \draw[-][line width=0.3mm][blue] (Drob) -- (Erob);
        \draw[-][line width=0.3mm][blue] (Erob) -- (Frob);
        \draw[-][line width=0.3mm][blue] (Frob) -- (Arob);
    
        \node at (-.5,3) {$u_{\mathcal{N}}(2)$};
        \node at (3,-.5) {$u_{\mathcal{N}}(1)$};
        \node[font=\small] at (2.6,1.4) {$F(\hat{\Xi}_N^1)$};
        \node[font=\small] at (2.4,4.8) {$F(\hat{\Xi}_N^2)$};
        \node[font=\small] at (3.1,3.1) {$A^{\beta}$};
        \node[font=\tiny] at (\pTwo+0.05,-0.3) {$p_{\hat{\Xi}_N^1}$};
        \node[font=\tiny] at (\pTworobust+0.05,-0.3) {$p_{\hat{\Xi}_N^2}$};
        \node[font=\tiny] at (\bOne+0.05,-0.3) {$b_{\hat{\Xi}_N^1}$};
        \node[font=\tiny] at (\bOnerobust+0.05,-0.3) {$b_{\hat{\Xi}_N^2}$};
        \draw[->][line width=0.4mm] (0,0) -- (6,0);
        \draw[->][line width=0.4mm] (0,0) -- (0,6);
        
\end{tikzpicture}
    \caption{A visualization of the proof of \cref{thm:robust_flex}. The robust set is the intersection of the aggregate flexibility sets generated by populations that lie in the ambiguity set. Here, we show the aggregate flexibility sets associated with two sample populations, $\hat{\Xi}_N^1$ and  $\hat{\Xi}_N^2$. The aggregate flexibility sets generated by them are, $F(\hat{\Xi}_N^1)$ and $F(\hat{\Xi}_N^2)$, and their intersection, $A^\beta$. The robust set is defined by the maximum of the supermodular functions (in this case $p_{\hat{\Xi}_N^2}$) and minimum of the submodular functions (in this case $b_{\hat{\Xi}_N^1}$). For clarity we only show the sub- and supermodular functions for $\mathcal{A} = \{1\}$.}
    \label{fig:visual_proof}
    \Description{A cube and a plank intersecting to form a g-polymatroid.}
  \end{figure}
Using this theorem we can characterize the robust aggregate flexibility set, $A^\beta$, as a g-polymatroid.
\begin{theorem}\label{thm:robust_flex}
    The robust aggregate flexibility set, $A^\beta$, is the g-polymatroid:
    \begin{equation}
        A^\beta = Q(p^\beta, b^\beta),
    \end{equation}
    where:
    \begin{align}\label{eq:robust_flex}
        p^\beta(\mathcal{A}) &:= \max_{\hat{\Xi}_N \in \mathbb{B}^\beta} p_{\hat{\Xi}_N}(\mathcal{A})\\
        b^\beta(\mathcal{A}) &:= \min_{\hat{\Xi}_N \in \mathbb{B}^\beta} b_{\hat{\Xi}_N}(\mathcal{A}).
    \end{align}
\end{theorem}
\begin{proof}

    For two g-polymatroids $Q(p_a, b_a)$ and $Q(p_b, b_b)$, the intersection of the two is a g-polymatroid $Q(p_c, b_c)$, where:
        \begin{align*}
            p_c(\mathcal{A}) &:= \max\{p_a(\mathcal{A}), p_b(\mathcal{A})\} \\
            b_c(\mathcal{A}) &:= \min\{b_a(\mathcal{A}), b_b(\mathcal{A})\} \quad \forall \mathcal{A} \subseteq \mathcal{T}.
        \end{align*}
    Evaluating the maximum of $p_{\hat{\Xi}_N}(\mathcal{A})$ and minimum of $b_{\hat{\Xi}_N}(\mathcal{A})$ over all elements in the ambiguity set $\mathbb{B}^\beta$ gives the required result.

\end{proof}
\cref{fig:visual_proof} provides a visualization of the proof of this theorem.

\section{Tractable Reformulation}\label{sec:reformulation}
Having introduced a theoretical framework for uncertainty quantification of aggregate flexibility sets, in this section we turn to the explicit computation of the parameters that define these sets.  Specifically, this section demonstrates how the optimization problems defining the supermodular function $p^\beta$ and submodular function $b^\beta$ from \cref{thm:robust_flex}, which together parameterize the g-polymatroid $Q(p^\beta, b^\beta)$, can be recast as finite convex programs. For brevity we will only focus on a reformulation of the supermodular function, $p^\beta$. Results for the reformulation of the submodular function can be found in the appendix.
We do this by considering, for a given subset $\mathcal{A} \subseteq \mathcal{T}$, the maximization problem
\begin{equation}
    p^\beta(\mathcal{A}) := \max_{\hat{\Xi}_N \in \mathbb{B}^\beta} p_{\hat{\Xi}_N}(\mathcal{A}), 
\end{equation}
where we are optimizing over empirical distributions $\mathbb{Q}_{\hat{\Xi}_N}$ that lie within $\mathbb{B}^\beta$, the Wasserstein ball of radius $\varepsilon$ centred at $\mathbb{P}$. For clarity in notation, we drop $\mathcal{A}$ in the notation of $p_{\hat{\Xi}_N}(\mathcal{A})$, denoting it simply as $p_{\hat{\Xi}_N}$.
Using the definition of $p_{\hat{\Xi}_N}$, we can write it in terms of the expectation of $p_{\xi}$ over the empirical distribution $\mathbb{Q}_{\hat{\Xi}_N}$:
\begin{equation}
    \frac{1}{N} p_{\hat{\Xi}_N} = \frac{1}{N} \sum_{\xi_i \in \hat{\Xi}_N} p_{\xi_i} =   \mathbb{E}_{\xi \sim \mathbb{Q}_{\hat{\Xi}_N}}[ p_{\xi}].
\end{equation}
And so we rewrite our optimization problem as a maximization of the expectation of $p_{\xi}$ with respect to the empirical distribution $\mathbb{Q}_{\hat{\Xi}_N}$:
\begin{equation}\label{eq:worst_case_expectation}
    p^\beta  = N \max_{\hat{\Xi}_N \in \mathbb{B}^\beta} \mathbb{E}_{\xi \sim \mathbb{Q}_{\hat{\Xi}_N}}[p_\xi]. \\
\end{equation}

As yet, we have not put any assumptions on the form of $\mathbb{P}$.
However, to make use of results in the literature, it is useful to have our Wasserstein ball centred on an empirical distribution.
Here we assume that $\mathbb{P}$ is given as an $M$-point empirical distribution, however this is not necessary as we discuss in \cref{appendix:projection}.
We write the support of this empirical distribution as $\Xi_{\mathbb{P}} = \{\xi_1, \ldots, \xi_M\}$.
\begin{definition}[Convexity of $p(\xi)$, $\Xi$]
    We say that the system of $p_\xi$ and $\Xi$ is convex if:
    \begin{itemize}
        \item[(1)] $\Xi \subseteq \mathbb{R}^d$ is a convex and closed set
        \item[(2)] $p_\xi$ can be written as the point-wise maximum of a finite number of affine concave functions, i.e.
    \begin{equation}
        p(\xi) := \max_{k\leq K} \{p_k(\xi)\},
    \end{equation}
    where $p_k(\xi)$ is an affine concave function, for all $k \leq K$.
    \end{itemize}
\end{definition}
From its definition in \eqref{eq:Xi_space}, $\Xi$ is a convex and closed set. 
From \cref{cor:individual_flex}, $p_\xi$ is clearly a point-wise maximum of a set of functions.
However, since it depends on the cardinality of the set $\mathcal{C}$, it is not concave. This is addressed in detail in \cref{appendix:convexity}.

From this, we can now invoke the following theorem and adapt it to suit our requirements.
\begin{theorem}[Worst-case distributions]\cite[Theorem 4.4]{MohajerinEsfahani2018Data-drivenReformulations}\label{thm:worst_case_reformulation}
    If $\;\Xi$ and $p(\xi)$ form a convex system, then the worst-case expectation \eqref{eq:worst_case_expectation} coincides with the optimal value of the finite convex program
    \begin{equation}
    \begin{aligned}
        p^\beta = &\max_{\alpha_{ik}, q_{ik}} \sum_{\xi_i \in \Xi_\mathbb{P}} \sum_{k=1}^K \alpha_{ik} p_k \left( \xi_i - \frac{q_{ik}}{\alpha_{ik}} \right) \\
        &\text{s.t.} \quad \frac{1}{M} \sum_{i=1}^M \sum_{k=1}^K \|q_{ik}\| \leq \varepsilon', \\
        &\quad \sum_{k=1}^K \alpha_{ik} = 1 \quad \forall i \leq M, \\
        &\quad \alpha_{ik} \geq 0 \quad \forall i \leq M, \; \forall k \leq K, \\
        &\quad \xi_i - \frac{q_{ik}}{\alpha_{ik}} \in \Xi \quad \forall \xi_i \in \Psi, \; \forall k \leq K.
    \end{aligned}
    \end{equation}
    \end{theorem}
\cref{thm:worst_case_reformulation} provides a tractable reformulation of the optimization problem \eqref{eq:worst_case_expectation}, allowing us to compute the worst-case expectation of the supermodular function $p^\beta$ efficiently.
The reformulation of the submodular function $b^\beta$ can be done in a similar manner, details of this can be found in \cref{appendix:submodular_reformulation}.


\section{Applications and Numerical Results}\label{sec:case_studies}
\begin{figure}[t!]
  \centering
  \includegraphics[width=0.464\textwidth]{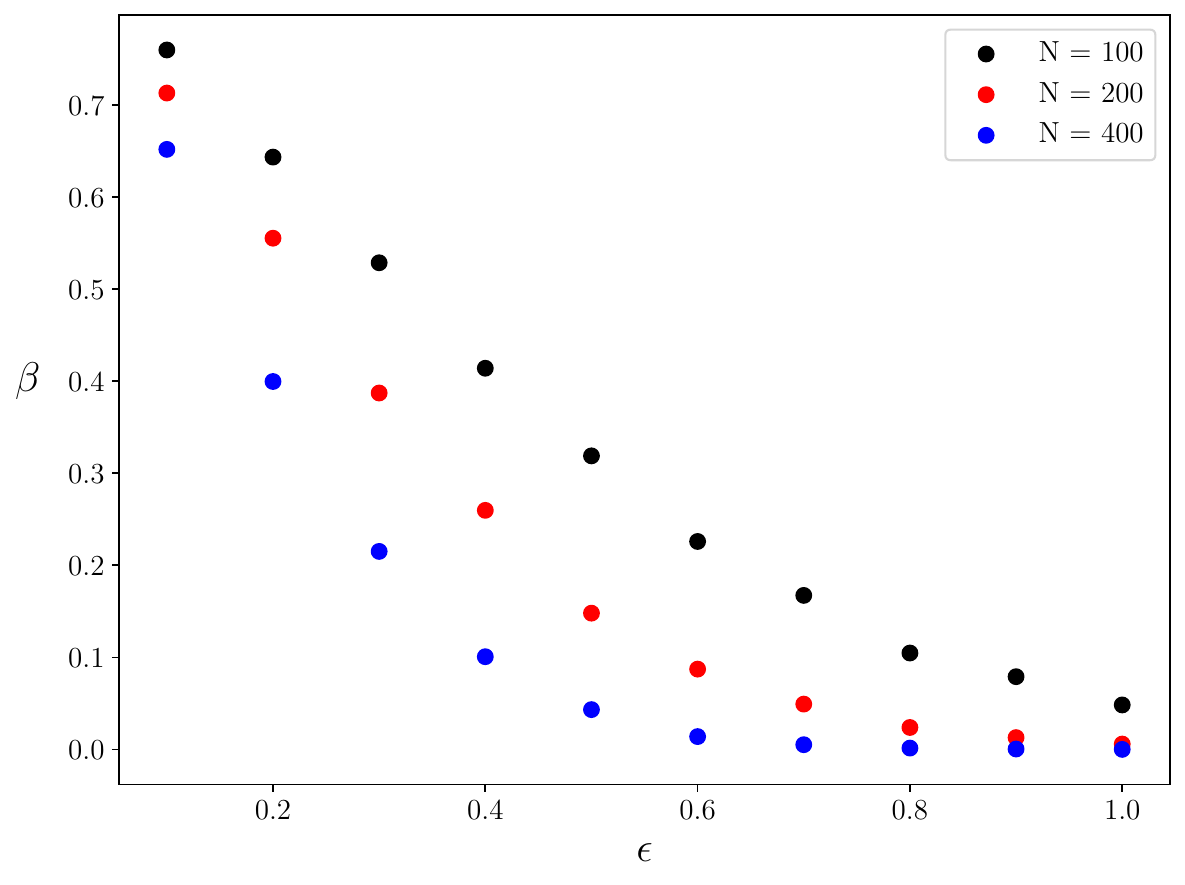}
  \caption{
  Empirical results showing the probability of \textit{not} being able to satisfy all aggregate charging profiles within the robust set
  $A^\beta$, for different values of $\varepsilon$ and $N$.
  }
  \label{fig:beta_epsilon}
\Description{A plot showing the probability of not being able to satisfy all aggregate charging profiles within $A^\beta$.}
\end{figure}
We now showcase the utility of the robust aggregation methods laid out in this work.
To use the results, we assume that an aggregator has access to historic data of the charging requirements:  hence, the empirical distribution generated by this data will make up $\mathbb{P}$. 
From $\mathbb{P}$ and a given size of EV population $N$, an aggregator wishes to quantify the confidence they have over the aggregate charging profiles that the population will be able to track. For example, they may be bidding into markets in which they are paid to track certain aggregate charging profiles, getting penalized if they deviate from the agreed charging profile as in \cite{Gade2024LeveragingMarkets}.  
Therefore, they may wish to quantify the confidence they have on the feasibility of tracking particular aggregate charging profiles, ideally providing tight bounds on this confidence. 
In this section, we provide some numerical results that showcase the theoretical results presented in this paper, and provide a simple case study to show how an aggregator might use these methods in practice. 

\subsubsection*{Validation of the theoretical results}
In this first study we showcase how the strong probabilistic guarantees of \cref{thm:measure_concentration} hold. 
To do so, we first synthesize a distribution $\mathbb{P}$ over charging requirements, uniformly sampling the space of charging requirements $\Xi$. Note that, in general, $\mathbb{P}$ could be any distribution over $\Xi$. 

Now, fixing $\varepsilon$ and  $N$, we construct the distributionally robust aggregate flexibility set  $A^\beta$, using \cref{thm:robust_flex} and its reformulation in \cref{thm:worst_case_reformulation}.
We then sample a population $\hat{\Xi}_N$ by drawing $N$ charging requirements from $\mathbb{P}$. 
Using \cref{cor:aggregate_flex}, we construct the aggregate flexibility set of the sampled population, namely $F(\hat{\Xi}_N)$. Finally, we check if $A^\beta$ is a subset of $F(\hat{\Xi}_N)$, i.e. ensuring that all aggregate charging profiles in $A^\beta$ are feasible for the sampled population.
We repeat this many times and measure the frequency that $A^\beta$ is contained in $F(\hat{\Xi}_N)$. From this frequency we can compute an empirical estimate for $\beta$. 
We generate aggregate flexibility sets for populations of EVs with various sizes, over a time horizon with $T=24$ steps. Constructing robust aggregate flexibility sets for different values of $\varepsilon$, we calculate $\beta$ for each of these sets. 
The guarantees of \cref{thm:measure_concentration} from \eqref{eq:beta_epsilon} establish an exponentially-decaying dependence of $\beta$ on the square of $\varepsilon$ — we can observe this relationship in \cref{fig:beta_epsilon}.
Similar results can be shown for the exponential decay of $\beta$ with respect to $N$.

\subsubsection*{Characterizing robust flexibility sets}
\begin{figure}[t!]
  \centering
  \raisebox{0mm}{
  \includegraphics[width=0.47\textwidth]{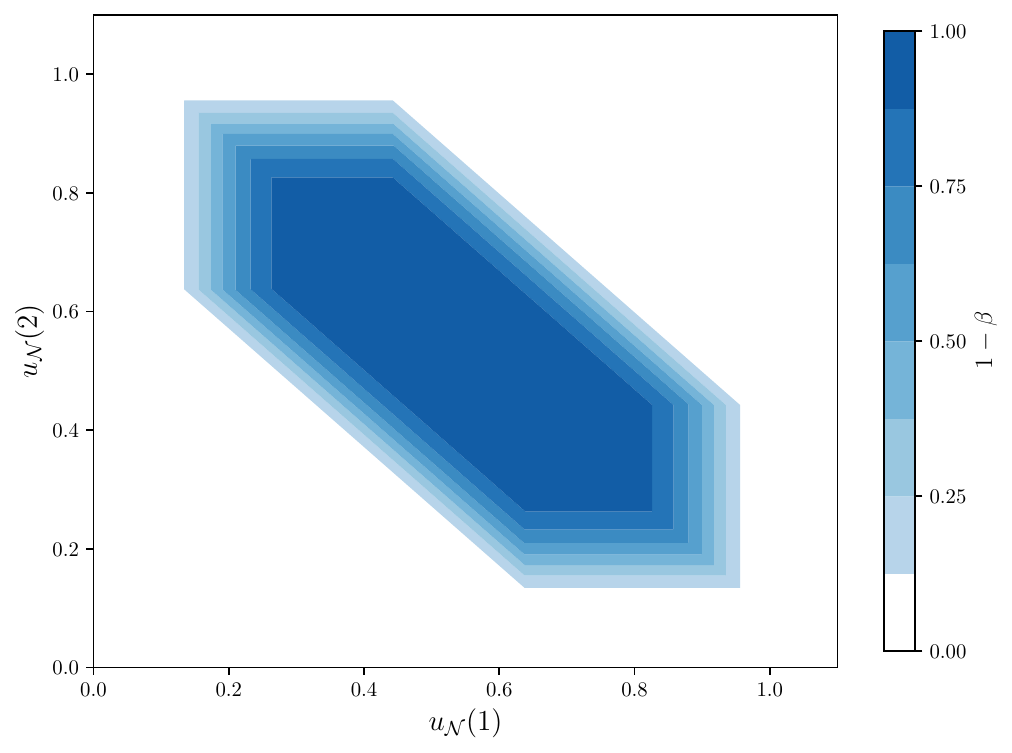}
  }
  \caption{
  Level sets of confidence for the robust aggregate flexibility sets $A^\beta$, over time horizon of $T=2$ steps.
  }
  \label{fig:robustness_region}
  
\Description{A plot showing the probability of not being able to satisfy all aggregate charging profiles within $A^\beta$.}
\end{figure}
In this second study, we show how the robust aggregate flexibility sets change with their confidence level, $\beta$. In order to be able to visualize the sets of interest, we consider the problem over a trivial time horizon of $T=2$ steps, such that $A^\beta \subset \mathbb{R}^2$.
Using the synthesized distribution $\mathbb{P}$ from the previous section, we construct the robust aggregate flexibility set, $A^\beta$, for a fixed population size of $N=100$, and for various values of $\beta$. We plot these robust aggregate flexibility sets in \cref{fig:robustness_region}, where different contours show the corresponding level sets of $\beta$.

Note that an almost identical plot can also be produced for the case where $\beta$ is fixed, and we vary $N$ to see how the robust aggregate flexibility sets change with the size of the population. 
By normalizing the robust aggregate flexibility sets, we can plot the flexibility allocated to each EV as a function of the population size $N$. This would illustrate how the confidence intervals per EV vary with different population sizes. Specifically, it highlights that larger populations achieve higher confidence levels per EV for the same set of charging profiles.
This motivates the utility of aggregators: with a given confidence level, larger populations of EVs allow for greater flexibility to be allocated to each individual EV.

\subsubsection*{Optimization over robust flexibility sets} 
\begin{figure}[t!]
  \centering
  \includegraphics[width=0.47\textwidth]{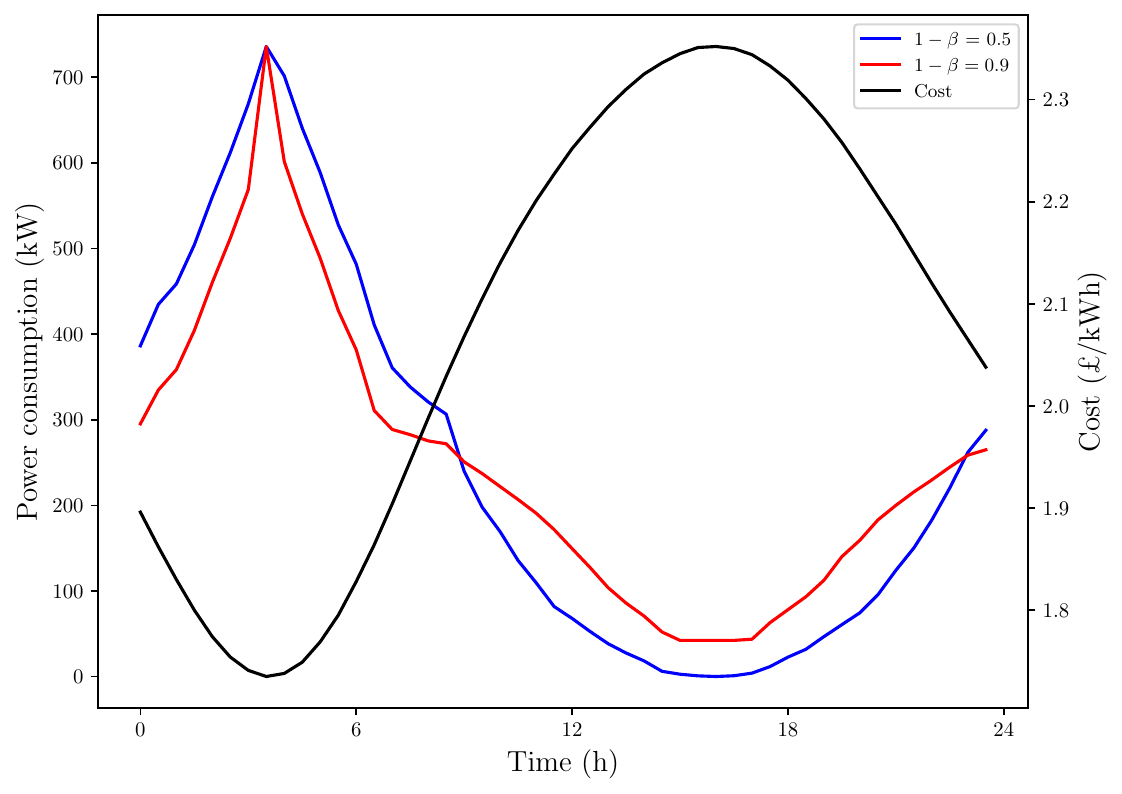}
  \caption{
  Aggregate charging profiles for different confidence levels when optimizing against a time-varying price signal.
  }
  \label{fig:robust_time_price}
\Description{A plot showing the probability of not being able to satisfy all aggregate charging profiles within $A^\beta$.}
\end{figure}

In this final study, we examine an aggregator bidding its EV fleet’s flexibility into a market subject to reliability constraints.
Such reliability requirements have recently been introduced by various system operators; for instance, the Danish transmission system operator mandates that ancillary service portfolios demonstrate at least 90\% availability \cite[10.2.2]{Energinet2024Prequalification2.1.2}. This directly maps onto setting $1 - \beta = 0.9$ in our framework. The aggregator's net cost $c(t)$ includes terms of the wholesale cost and revenues from dispatch in the reserve markets: $c(t) = c_w(t) - r_a(t)$. Formally, this leads to solving:
\begin{equation}
    \begin{aligned}
        \min_{u_{\mathcal{N}} \in \mathbb{R}^T} \quad & \sum_{t=1}^T c(t) u_{\mathcal{N}}(t) \\
        \text{s.t.} \quad & u_{\mathcal{N}} \in  A^\beta.
    \end{aligned}
\end{equation} 
For this case study we consider a population of $N=100$ EVs, and a time horizon of $T=48$ half-hourly time periods. We generate the distribution $\mathbb{P}$ by sampling $N$ charging requirements from a uniform distribution the parameter space $\Xi$. We construct the robust aggregate flexibility set, $A^\beta$, for various values of $\beta$. We then solve the optimization problem for each of these sets, and calculate the cost of the optimal charging profile, showing the outcomes in \cref{fig:robust_time_price}. 

When choosing the charging profile with higher certainty ($1-\beta = 0.9$), the aggregator is less able to exploit the cheaper energy in the early hours of the day and consumes more energy in the evening when prices are higher, as they must be conservative in their estimation of the aggregate flexibility set.
Clearly, the cost of charging increases as the required probability of satisfying the aggregate charging profile increases. This is to be expected, as the aggregator is forced to be more conservative over their estimation of aggregate flexibility sets as the confidence level increases. 
\section{Implications and Future Work}\label{sec:conc}

Extending previous work that gives exact characterizations of aggregate flexibility sets, we have shown how individual flexibility sets of EVs are members of a family of polytope known as generalized polymatroids.
Computing the Minkowski sum of g-polymatroids is efficient, allowing us to provide a tractable method for computing the exact aggregate flexibility set for populations of EVs.
Building on this we assume EV charging requirements are uncertain and i.i.d. according to a known distribution.
We exploit theoretical results that enable us to characterize robust aggregate flexibility sets, sets of aggregate charging profiles that can be tracked with a given confidence level.
The theoretical results leverage powerful finite-sample guarantees that enable us to provide tight bounds on this confidence.
We show explicitly how to compute the variables that parameterize these sets.
Finally, we demonstrate the soundness of the characterizations by means of numerical experiments and show how an aggregator might use these methods in practice.

This work can also be used to further motivate the utility of aggregators. As entities that aggregate the uncertainty of individual EVs, they can unlock more flexibility in the populations they control, than would be possible with individual EVs alone. 

Future work will focus on extending the results to more general charging models. We could consider the case where EVs are allowed to discharge energy back into the grid. Furthermore, we could consider the case where the charging requirements of the EVs are not i.i.d.. Maintaining the independence assumption, but allowing EVs to have different distributions over their charging requirements, would be a natural and realistic extension. Additionally, there are various power systems problems that require uncertainty quantification techniques when aggregating EV flexibility. One could, for example, consider newsvendor-like problems, where aggregators bid the flexibility of uncertain populations of EVs into flexibility markets — this paper provides a theoretical framework for such applications. 

\section*{Acknowledgments} 
The authors are grateful to Georg Loho for insightful guidance on g-polymatroids, to Yannik Schnitzer for his comments on earlier drafts, and to Clara Dijkstra for her helpful contributions to one of the proofs.
The work of C. Qu is supported in part from NSFC through 723B1001. The work of P. You is supported in part from NSFC through 723B1001, 72431001, 72201007, T2121002, 72131001.

\bibliographystyle{ACM-Reference-Format}
\bibliography{references}
\appendix
\section{Reformulation of Optimization}\label{appendix:reformulation}
\subsection{Projection onto empirical distribution}\label{appendix:projection}
To avoid introducing any unnecessary assumptions about $\mathbb{P}$, we can center our ball on the projection (via the Wasserstein distance) of $\mathbb{P}$ onto the set of $M$-point empirical distributions, and enlarge the radius of our ambiguity set.
Specifically, we use the empirical distribution that minimizes the Wasserstein distance:
\begin{equation}
    \mathbb{P}_N := \underset{\mathbb{P}_M \in \mathcal{P}_M(\Xi)}{\textrm{argmin}} \;\;d_W(\mathbb{P}_M, \mathbb{P}),
\end{equation}
as the center of our ambiguity set. Moreover, taking the upper bound of the triangle inequality,
\begin{equation}
    d_W(\mathbb{P}_M, \mathbb{Q}_M) \leq  d_W(\mathbb{P}_M, \mathbb{P}) +  d_W(\mathbb{P}, \mathbb{Q}_M),
\end{equation}
 we update the radius of our ambiguity set with 
 \begin{equation}
        \varepsilon_M(\beta)' = d_W(\mathbb{P}_M, \mathbb{P}) + \varepsilon_M(\beta),
 \end{equation}
 to ensure the probabilistic guarantees of \cref{thm:measure_concentration} still hold. 
 There exists a theoretical bound, $d_W(\mathbb{P}_M, \mathbb{P}) \leq C M^{-1/d}$, where $d$ is the dimension of the support of $\mathbb{P}$ and $C$ is a constant that depends on the distribution $\mathbb{P}$.
 However, in practice, one can evaluate $d_W(\mathbb{P}_M, \mathbb{P})$ exactly, once $\mathbb{P}_M$ has been computed.
 
 Furthermore, $\mathbb{P}$ will most likely be made up of historical data and as such will already be an empirical distribution, in which case we simply use this and leave the radius of the ambiguity set unchanged.
\subsection{Convexity of sub- and supermodular functions}\label{appendix:convexity}
In their definitions $p_\xi$ and $b_\xi$ take $|\mathcal{C}|$, a set of discrete values, as an argument, and so these functions are not affine. However we can show that they are bounded by the following:
\begin{align*}
    &p_\xi(\mathcal{A})  \geq \textrm{max}_{j,k} \{0, \underline{e} - T_{jk}m\} \\
    &b_\xi(\mathcal{A})  \leq \textrm{min}_{j,k} \{T_{jk}m, \overline{e}\}
\end{align*}
where the equality holds when $\underline{t}, \overline{t} \in \mathbb{Z}$, and we define:
\begin{equation}
    T_{jk} := (\underline{a}_j \underline{t} + \overline{a}_k \overline{t}) + \underline{b}_j + \overline{b}_k
\end{equation}
and $\underline{a}_j, \overline{a}_k, \underline{b}_j, \overline{b}_k$ are constants are depend on $\mathcal{A} \subseteq \mathcal{T}$. Taking the upper bound on $p_\xi(\mathcal{A})$ and the lower bound on $b_\xi(\mathcal{A})$ we can write $p_\xi$ and $b_\xi$ as the point-wise maximum and minimum of a finite number of affine concave functions, respectively.

\subsection{Reformulation of worst expectation of submodular function}\label{appendix:submodular_reformulation}
In \cref{sec:reformulation} we showed how the robust aggregate flexibility sets can be reformulated into finite convex optimization problems. Here we provide the full details of the reformulation of the submodular function, $b^\beta$, defining $A^\beta = Q(p^\beta, b^\beta)$. Viewing the submodular function as the negative of a supermodular function we can use the same form for the reformulated optimization problem as the supermodular function. Therefore, we simply write the reformulation as a maximization over its negative:
\begin{equation}
    \begin{aligned}
        b^\beta = - &\max_{\alpha_{ik}, q_{ik}} - \sum_{\xi_i \in \Xi_\mathbb{P}} \sum_{k=1}^K \alpha_{ik} b_k \left( \xi_i - \frac{q_{ik}}{\alpha_{ik}} \right) \\
        &\text{s.t.} \quad \frac{1}{M} \sum_{i=1}^M \sum_{k=1}^K \|q_{ik}\| \leq \varepsilon', \\
        &\quad \sum_{k=1}^K \alpha_{ik} = 1 \quad \forall i \leq M, \\
        &\quad \alpha_{ik} \geq 0 \quad \forall i \leq M, \; \forall k \leq K, \\
        &\quad \xi_i - \frac{q_{ik}}{\alpha_{ik}} \in \Xi \quad \forall \xi_i \in \Psi, \; \forall k \leq K.
    \end{aligned}
    \end{equation}

From which we can explicitly calculate $b^\beta$, can complete our characterization of the robust aggregate flexibility sets.

\end{document}